\documentclass[aps,prl,preprint,superscriptaddress,showpacs,amsmath,amssymb,lineno]{revtex4-1}

\usepackage{graphicx}
\usepackage{dcolumn}
\usepackage{bm}

\begin{document}

\title{Noise-Induced Synchronization of a Large Population of Globally Coupled Nonidentical Oscillators}
\author{Ken H. Nagai}
\email[E-mail: ]{nagai.ken@ocha.ac.jp}
\affiliation{Division of Advanced Sciences, Ochadai Academic Production, Ochanomizu University, Tokyo, 112-8610, Japan}
\author{Hiroshi Kori}
\affiliation{Division of Advanced Sciences, Ochadai Academic Production, Ochanomizu University, Tokyo, 112-8610, Japan}
\affiliation{PRESTO, Japan Science and Technology Agency, Kawaguchi, Saitama, 332-0012, Japan}
\date{\today}

\begin{abstract}
We study
a large population of globally coupled phase oscillators subject to
common white Gaussian noise and find analytically that the critical
coupling strength between oscillators for synchronization
transition decreases with an increase in the
intensity of common noise.
Thus, common noise promotes the onset of synchronization. 
Our prediction is confirmed by numerical
simulations of the phase oscillators as well as of limit-cycle oscillators.
\end{abstract}

\pacs{05.45.Xt, 05.10.Gg, 05.40.Ca, 05.70.Fh}

\maketitle 

Synchronization of an
ensemble of periodic oscillators has attracted considerable attention
because of its broad applications in many fields ranging from physics to
engineering~\cite{Kuramoto1984, Winfree2001, KissRusinKoriHudson2007, Ermentrout1996, *MikhailovShowalter2006, *EckhardtOttStrogatzAbramsMcRobie2007}.
In particular, synchronization plays an essential role in numerous
biological functions,
including the formation of pacemaker tissues of the heart and of the
circadian master clock \cite{Glass2001, *ReppertWeaver2002}.

Because real systems are inevitably subject to noise, it is important to
understand the effect of noise on the synchronization of periodic
oscillators. Some types of noise, including thermal noise or intrinsic
noise in cells, act independently on individual components, which
usually inhibits
synchronization~\cite{Kuramoto1984,Sakaguchi1987,*StrogatzMirollo1991}.
However, there are many situations where a single noise process, such as
that originating from environmental fluctuations, acts on an entire
system.  Whether such common noise enhances or inhibits synchronization
is actually unclear.  This issue is thought to be relevant to biological
pacemaker tissues in that external noise could have a positive effect on
synchronization. However, clarification of the outcome of fluctuating
input is necessary in cases, such as that of deep brain stimulation for
Parkinson's disease~\cite{Tass2003}, in which global external
stimulation is used to destroy synchronization of dynamic components.

The effect of common noise on uncoupled oscillators or
coupled-oscillator networks with small sizes
has been
extensively studied for both periodic and chaotic oscillators,
and rigorous theoretical frameworks have been proposed~\cite{Pikovskii1984, *PikovskyRosenblumKurths2001, *UchidaMcAllisterRoy2004,
*TeramaeTanaka2004, *NagaiNakaoTsubo2005, *NakaoAraiKawamura2007, *LyErmentrout2009}.
In contrast, for a large population of coupled oscillators,
despite a numerous body of numerical and experimental evidence
\cite{[{see, e.g.,}] [{}]ZhouKurthsKissHudson2002, *ParkLaiKrishnamoorthyKandangath2007, *Sakaguchi2008, *GilKuramotoMikhailov2009}, 
theoretical treatment is still an open and challenging problem.


In this letter, we 
investigate a large population of nonidentical phase oscillators
that are globally coupled and subject
to common Gaussian white noise to clarify the
effect of common noise on coupled oscillators.  Utilizing the
anzatz recently proposed by Ott and
Antonsen~\cite{OttAntonsen2008,*OttAntonsen2009}, 
we analytically show
that the addition of common noise leads to a decrease in the critical
coupling strength for synchronization transition. Our prediction is
corroborated by direct numerical simulations of the model.  We also
numerically confirm that globally coupled limit-cycle oscillators show
the same dependence on common noise.  
The employed phase model approximates many realistic systems
with weak coupling and weak forcing. Thus, our results suggest that
weak common noise generally promotes synchronization of oscillators with
week and global coupling.

Consider 
globally coupled phase oscillators, 
known as the Sakaguchi-Kuramoto model \cite{SakaguchiKuramoto1986}, subject to a
common external force
\begin{equation}
\frac{{\rm d}\theta_{i}}{{\rm d}t} = 
 \omega_{i}+\frac{K}{N}\sum_{j=1}^{N} 
 \sin(\theta_{j}-\theta_{i}+\beta)+  p(t) \sin \theta_{i},
 \label{model}
\end{equation}
where $\theta_i$ and $\omega_i$ are the phase and the natural frequency,
respectively, of the $i$-th oscillator, $K>0$ is the coupling strength,
$\beta$ is a parameter of the coupling function ($-\pi/2<\beta<\pi/2$), and $p(t)$ is a
common external force.  We assume that the natural frequency distribution is
given by a Lorentzian function $f_{\rm
freq}(\omega) = \frac{1}{\pi}\frac{1}{(\omega-\omega_{0})^{2}+1}$.
We will 
further assume white Gaussian noise for $p(t)$~\footnote{As pointed out
in~\cite{TeramaeNakaoErmentrout2009,*YoshimuraArai2008} an additional term may appear in the phase model for the case with white noise. However, in our analysis, such a term does not change the transition behavior, so we neglect it.},
but first we treat $p(t)$ as a general time-dependent function for a
while.

Note that Eq. (\ref{model}) approximates
various realistic oscillators with weak coupling and weak forcing
\cite{Kuramoto1984,KissRusinKoriHudson2007,KoriKawamuraNakaoAraiKuramoto2009}.
Note also that the common external force is
multiplied by a function of the phase, $\sin \theta_i$, which is called
a phase sensitivity function. The phase sensitivity function naturally
appears in the phase description of limit-cycle
oscillators~\cite{Kuramoto1984, Winfree2001,
KoriKawamuraNakaoAraiKuramoto2009}.
We will later demonstrate these facts using a limit-cycle-oscillator model
that generally appears near a Hopf bifurcation.


We examine 
the synchronization transition in the large-$N$ limit. For a better presentation, we
set $\beta=0$ (corresponding to the Kuramoto model
\cite{Kuramoto1984}). The extension to nonzero $\beta$ is straightforward;
we will only show a final result for nonzero $\beta$ in the present
paper. In the limit $N\to \infty$, Eq.~\eqref{model}
becomes
\begin{equation}
\frac{\partial f}{\partial t}+ \frac{\partial}{\partial\theta}\Bigg\{\Bigg(\omega+K\frac{re^{-{\rm i}\theta}
 -r^{\ast}e^{{\rm i}\theta}}{2{\rm i}}+
 \frac{e^{{\rm i}\theta}-e^{-{\rm i}\theta}}{2{\rm i}}
 p \Bigg)f\Bigg\}=0,\label{phase_distribution}
\end{equation}
where $f(\omega,\theta,t)$ is the distribution function for the phases
of the oscillators with natural frequency $\omega$,
$r=\int_{-\infty}^{\infty} \! {\rm d}\omega\int_{0}^{2\pi} \! {\rm
d}\theta \ fe^{{\rm i}\theta}$ is the Kuramoto order parameter, and
$\ast$ represents the complex conjugate.
For $p(t)=0$, the synchronization transition (the so-called Kuramoto
transition) occurs at $K=K_{\rm c}= 2$, above which $|r|$ is
nonvanishing~\cite{Kuramoto1984}.

To investigate the transition in Eq.~\eqref{phase_distribution}, we first
derive a dynamical equation for the order parameter $r$. For this, we
employ the Ott-Antonsen ansatz for the phase distribution
\begin{equation}
f=\frac{f_{\rm freq}(\omega)}{2\pi}\left\{1+\sum_{n=1}^{\infty}\left[\left(\alpha e^{{\rm i}\theta}\right)^{n}+\left(\alpha^{\ast}e^{-{\rm i}\theta}\right)^{n}\right]\right\},\label{anzatz}
\end{equation}
where $\alpha(\omega, t)$ is a certain function~\cite{OttAntonsen2008, *OttAntonsen2009}.
Substituting Eq.~\eqref{anzatz} into Eq.~\eqref{phase_distribution}, we
obtain a dynamical equation for $\alpha$
\begin{equation}
\frac{\partial\alpha}{\partial t}+\frac{K}{2}(r\alpha^{2}-r^{\ast})+{\rm i}\omega\alpha+\frac{p}{2}(1-\alpha^{2})=0.\label{alpha_eq}
\end{equation}
Note that $r(t)=\alpha^{\ast}(\omega_{0}-{\rm i},t)$ because
$r=\int_{-\infty}^{\infty} {\rm d}\omega\int_{0}^{2\pi} {\rm d} \theta f
e^{{\rm i}\theta}=\int_{-\infty}^{\infty} {\rm d}\omega f_{\rm freq}\alpha^{\ast}$ and $f_{\rm freq}(\omega)= \frac{1}{\pi}\frac{1}{(\omega-\omega_{0}+{\rm i})(\omega-\omega_{0}-{\rm i})}$.
Thus, by setting $\omega=\omega_{0}-{\rm i}$ in
Eq.~\eqref{alpha_eq}, we obtain
\begin{equation}
\frac{{\rm d}r}{{\rm d}t}=\left(-1+\frac{K}{2}+{\rm i}\omega_{0}\right)r-\frac{K}{2}|r|^{2}r-\frac{p}{2}(1-r^{2}).
\end{equation}
By letting $r=\sqrt{A}e^{{\rm i}(\omega_{0}t+\phi)}$,
we further obtain
\begin{align}
\frac{{\rm d}A}{{\rm d}t}&=h(A)+g_{A}(A,\omega_{0}t+\phi)p,\\
\frac{{\rm d}\phi}{{\rm d}t}&=g_{\phi}(A,\omega_{0}t+\phi)p, 
\label{order_dynamics}
\end{align}
where $h(A)=\left(K-2\right)A-KA^{2}$,
$g_{A}(A,\omega_{0}t+\phi)=-\sqrt{A}(1-A)\cos(\omega_{0}t+\phi)$,
and $g_{\phi}(A,\omega_{0}t+\phi)=\frac{(1+A)}{2\sqrt{A}}\sin(\omega_{0}t+\phi)$.

Now we assume that $p(t)$ is white Gaussian noise
with $\left<p(t)\right>=0$ and $\left<p(t)p(s)\right>=2D\delta(t-s)$, and 
interpret Eq.~\eqref{order_dynamics} as a Stratonovich differential equation.
Then we obtain the Fokker-Planck equation for the probability
distribution $q(A,\phi,t)$, given by
\begin{align}
\frac{\partial q}{\partial t}=-\frac{\partial}{\partial
 A}\left\{\left(h+D\left(g_{A}\frac{\partial g_{A}}{\partial
 A}+g_{\phi}\frac{\partial g_{A}}{\partial
 \phi}\right)\right)q\right\}-\frac{\partial}{\partial\phi}\left\{D\left(g_{\phi}\frac{\partial
 g_{\phi}}{\partial\phi}+g_{A}\frac{\partial g_{\phi}}{\partial
 A}\right)q\right\}\notag\\
+D\left(\frac{\partial^{2}}{\partial
 A^{2}}\left(g_{A}^{2}q \right)+2\frac{\partial^{2}}{\partial A\partial\phi}\left(g_{A}g_{\phi}q\right)+\frac{\partial^{2}}{\partial\phi^{2}}\left(g_{\phi}^{2}q\right)\right).\label{FP-eq1}
\end{align}
Because $h$, $g_{A}$, $g_{\phi}$, and $q$ are $2\pi$--periodic
functions, integrating of both sides of Eq.~\eqref{FP-eq1} over
$\phi$ from $0$ to $2\pi$ yields
\begin{align}
\frac{\partial Q}{\partial t}&=-\frac{\partial}{\partial
 A}\left\{\int^{2\pi}_{0}\left(h+D\left(g_{A}\frac{\partial
 g_{A}}{\partial A}+g_{\phi}\frac{\partial g_{A}}{\partial
 \phi}\right)\right)q \ {\rm d}\phi\right\}
 +D\frac{\partial^{2}}{\partial A^{2}}\left(\int^{2\pi}_{0}g_{A}^{2}q \
 {\rm d}\phi\right)\label{FP-eq2}
\end{align}
where $Q(A,t)=\int_{0}^{2\pi} q \ {\rm d}\phi$.

At this stage, we additionally assume that $K$ and $D$ are sufficiently
small compared to a typical natural frequency $\omega_{0}$. It is
natural to assume this because
this is the condition under which
Eq.~\eqref{model} approximates coupled limit-cycle oscillators.
Under this assumption, $Q$ evolves sufficiently slowly compared to a typical
oscillation time scale, i.e., $2\pi/\omega_{0}$. Thus, to a good approximation,
the right-hand side of Eq.~\eqref{FP-eq2} can be time-averaged over the
duration of $2\pi/\omega_{0}$, leading to
\begin{equation}
\frac{\partial Q}{\partial t}=-\frac{\partial}{\partial
 A}\left\{\left(\frac{D}{2}+\left(K-2-D\right)A -\left(K-\frac{D}{2}\right)A^{2}\right)Q\right\}
+\frac{\partial^{2}}{\partial A^{2}}\left\{\frac{D}{2}A(1-A)^{2}Q\right\}.
\end{equation}
Letting ${\partial Q}/{\partial t}=0$, we obtain the stationary
distribution $Q_{\infty}(A)$ as
\begin{align}
Q_{\infty}(A)
&=C\exp\left[\frac{2}{D}\left\{-\frac{2A}{1-A}-(K+D)\log(1-A)\right\}\right],\label{bunpu}
\end{align}
where $C=1/\int_{0}^{1}\!{\rm d}A \exp\left[\frac{2}{D}\left\{-\frac{2A}{1-A}-(K+D)\log(1-A)\right\}\right]$.
In stochastic systems, the maximum of the probability distribution
function is often adopted as the order parameter characterizing a
transition~\cite{HorsthemkeLefever1983}.  From
Eq. (\ref{bunpu}), it follows that $Q_{\infty}(A)$ assumes its maximum
at
\begin{equation}
A_{\rm max}=
\begin{cases}
0 & (K+D<2)\\
\frac{K+D-2}{K+D} & (K+D\geq 2)
\end{cases}
.\label{Amax}
\end{equation}
Thus we find that the critical coupling strength at which $A_{\rm
max}$ becomes nonvanishing is $K_{\rm
c}=2-D$; the common noise decreases the critical
coupling strength by $D$ as compared to that in the original Kuramoto transition.

For nonzero $\beta$, one can show that $K$ in Eq. (\ref{Amax})
is replaced by $K \cos \beta$. Thus, the critical condition is given by $K_{\rm c}=\frac{2-D}{\cos\beta}$.

\begin{figure}
\includegraphics{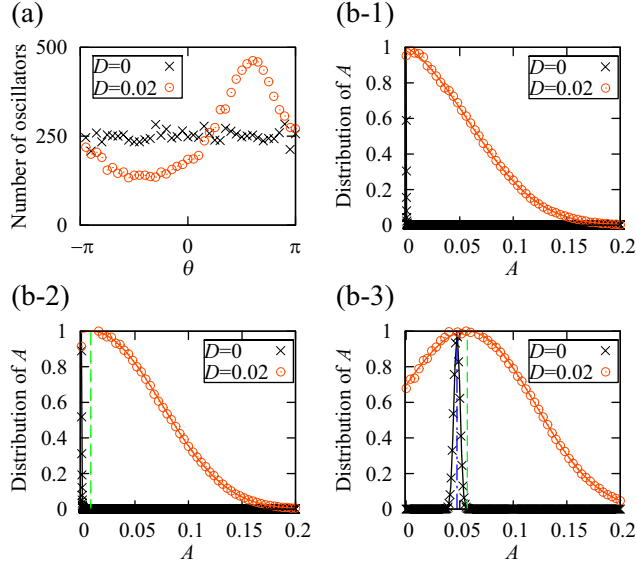}
\caption{(color online) Numerical results for the phase model given by
 Eq.~(\ref{model}). Crosses (black) and open circles (orange) represent 
 numerical data for $D=0$ and $D=0.02$, respectively.  (a) Snapshot of
 the phase distribution for $K=1.99$. (b) Distribution of $A$ for (b-1)  $K=1.96$, (b-2) $K=1.99$, and (b-3) $K=2.1$. Lines on the points are
 fitting curves. Histograms and curves are normalized for the maximum of curves to be 1. Point-dashed line (blue) and dashed line (green)
 represent the numerically identified $A_{\rm max}$ for $D=0$ and $D=0.02$,
 respectively.}
\label{order}
\end{figure}

\begin{figure}
\includegraphics{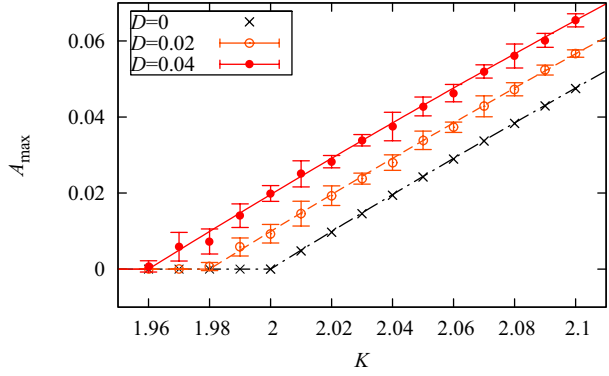}
\caption{(color online) $A_{\rm max}$ for the
 phase model as a function of $K$. Crosses (black), open circles (orange)
 and filled circles (red) indicate the numerically identified
 $A_{\rm max}$ for $D=0$, $D=0.02$, and $D=0.04$, respectively.  Error
 bars represent the variance of $A_{\rm max}$ for 10 trials with different initial conditions and different noise
 processes. Point-dashed line (black), dashed line (orange), and a
 solid line (red) represent  Eq.~\eqref{Amax} for $D=0$, $D=0.02$, and $D=0.04$, respectively.}
\label{phase_d}
\end{figure}

We confirmed our prediction by numerical simulation of
Eq.~\eqref{model} with $N=10000$ and $\beta=0$.
The Lorentzian distribution for the natural frequency was given by~\cite{Daido1986}
\begin{equation}
\omega_{i}=\omega_{0}+\tan\left\{i\frac{\pi}{N}-(N+1)\frac{\pi}{2N}\right\}\quad(1\leq
 i\leq N). \label{omega_bunpu}
\end{equation}
We set $\omega_{0}=100$ to ensure that $K$ and $D$ are much smaller than
$\omega_{0}$.  We employed random initial conditions and numerical data
were obtained from $t=10000$ to $t=60000$.  As shown in Fig.~\ref{order}
(a), the phase distribution did not cluster for $K=1.99$ and $D=0$ $(K+D<2)$. In contrast, a cluster
of oscillators was observed for $K=1.99$ and $D=0.02$ $(K+D>2)$.
To estimate $A_{\rm max}$ from the numerical data, the logarithm of the
histogram of $A$ around the peak was fitted to the
logarithm of Eq.~\eqref{bunpu}, i.e., $a+\frac{2}{b}\left(-\frac{2A}{1-A}-(c+b)\log(1-A)\right)$ with fitting parameters $a$, $b$, and $c$.
The obtained data were well fitted [Fig.~\ref{order} (b)].
The numerically identified values of $A_{\rm max}=\frac{b+c-2}{b+c}$
were plotted in Fig.~\ref{phase_d}, which shows excellent agreement with
the theoretical prediction of Eq.~\eqref{Amax}.
In our preliminary numerical simulations, we also confirmed that a
similar transition behavior occurs in the case of the Gaussian
distribution for the natural frequency (data not shown).

\begin{figure}
\includegraphics{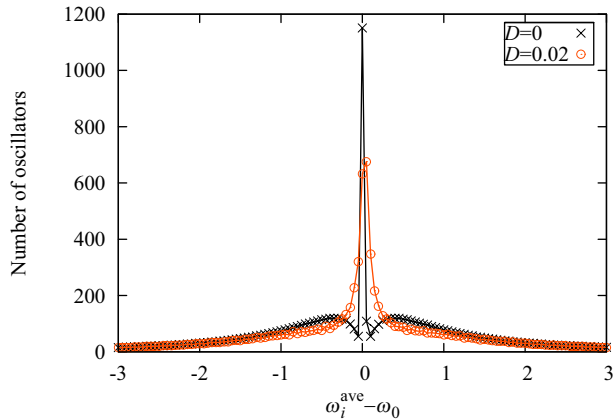}
\caption{(color online) Distribution of the long-time averaged
 frequencies of the phase oscillators for $K=2.02$. Crosses (black) and open circles (orange) with connecting lines are the numerical results for $D=0$ and $D=0.02$, respectively.}
\label{omegabunpu}
\end{figure}

We also observed the
distribution of the averaged frequencies $\omega_{i}^{\rm ave}$
defined as the long-time average of $\dot{\theta}_{i}$. Numerical
results are shown in Fig.~\ref{omegabunpu}. Without noise, the
distribution had a delta-function peak at $\omega_{0}$, whereas for $D \neq 0$, this peak disappeared and
the distribution was continuous.  This qualitative difference can be
explained as follows. Using the Kuramoto order parameter $r$,
Eq.~\eqref{model} can be written as $\dot \theta_{i} =  \omega_{i}+K |r|
\sin(\omega_{0} t+\phi -\theta_{i})+  p(t) \sin \theta_{i}$.
For $D=0$, $|r|$ and $\phi$ are time-independent
after transient~\cite{Kuramoto1984}. Then, oscillators with
$|\omega_{i}-\omega_{0}|< |r|$ are phase-locked to the mean field, so
that their actual frequencies are exactly the same as
that of the mean field, which is $\omega_0$.
However, for $D \neq 0$, 
$|r|$ fluctuates with time and becomes vanishingly small with
a finite probability [see Eq.~\eqref{bunpu} and Fig.~\ref{order} (b)].
This
implies that any oscillator except that with $\omega_{i}=\omega_0$
cannot be phase-locked to the mean field for an infinitely long time.
Therefore, oscillators with $\omega_{i} > \omega_{0}$ ($\omega_i <
\omega_0$) tend to have a larger (smaller) averaged frequency
than that for $D=0$, so that the delta-function peak vanishes.

Finally, we demonstrate the validity of our prediction in limit-cycle
oscillators. 
We introduce the following model
\begin{equation}
\frac{{\rm d}W_{i}}{{\rm d}t}=(1+{\rm i}\omega_{i})W_{i}-\left|W_{i}\right|^{2}W_{i}+\frac{\epsilon K}{N}\sum_{j=1}^{N}(W_{j}-W_{i})+\sqrt{\epsilon}p(t),\label{sl}
\end{equation}
where $W_i$ is the complex state variable of the $i$-th oscillator,
$\epsilon$ is a small parameter to denote that the coupling strength and the noise
strength smaller than both the relaxation rate of the amplitude dynamics
and the natural frequencies of oscillators, and
$p(t)$ is a common white Gaussian noise with strength $D$.  Each individual oscillator is
called a Stuart-Landau oscillator, which generically appears when the
system is near a Hopf bifurcation~\cite{Kuramoto1984}. Eq.~\eqref{sl} is
approximated by Eq.~\eqref{model} with $\beta=0$ for small
$\epsilon$~\cite{Kuramoto1984}, so 
similar behavior is expected.

\begin{figure}
\includegraphics{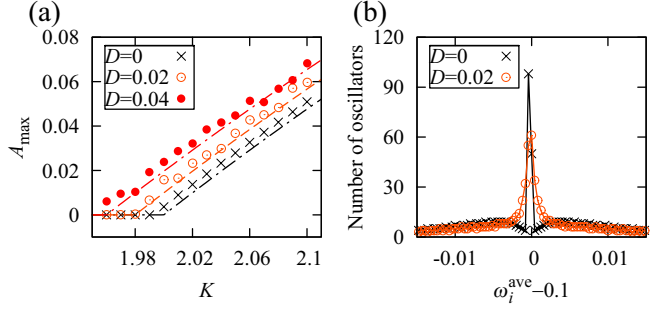}
\caption{(color online) Numerical results for the limit-cycle
oscillators given by Eq. (\ref{sl}). Legends for (a) and (b) are the
 same as those in Figs.~\ref{phase_d} and \ref{st_A}, respectively.
(a) Order parameter $A_{\rm max}$ as a function of $K$. Lines represent Eq.~\eqref{Amax}. 
(b) Distribution of the long-time averaged frequencies for $K=2.02$. We
used the data from $t=5\times 10^{6}$ to $t=10\times
10^{6}$. $N=1000$, $\epsilon=0.01$, and
$\omega_{i}=0.1+\epsilon\tan\left\{i\frac{\pi}{N}-(N+1)\frac{\pi}{2N}\right\}$.
}
\label{st_A}
\end{figure}

We numerically simulated Eq.~\eqref{sl} with $N=1000$.  We defined $A$ as
$|\sum_{j=1}^{N}e^{{\rm i}\theta_{j}}/N |^{2}$ with $\theta_j = \arg
W_{j}$ and estimated $A_{\rm max}$ in the same manner as for
the phase oscillators. The numerically determined $A_{\rm max}$ values are shown
in Fig.~\ref{st_A} (a), which agrees reasonably well with the prediction
of Eq.~\eqref{Amax}. We also observed that the distribution of
$\omega_{i}^{\rm ave}$ was continuous for $D\neq 0$ [Fig.~\ref{st_A} (b)].

To conclude, we have studied the Sakaguchi--Kuramoto model subject to common
noise and analytically showed that the critical coupling strength for
the synchronization--desynchronization transition decreases with an
increase in the strength of the common noise. The prediction has been
numerically corroborated. We have also found that the distribution of
the averaged frequencies is continuous when common noise is present.
Our results suggest that weak common noise generally promotes
synchronization of weakly coupled oscillators.
It would be interesting to experimentally investigate
the effect of common noise on coupled biological and chemical oscillators.

We thank Dan Tanaka for motivating us to study this topic.
We also thank Hayato Chiba and Hiroya Nakao for helpful discussions.

%

\end{document}